# The quantum dynamics of two coupled large spins

V.E. Zobov

*L.V.Kirensky Institute of Physics, Russian Academy of Sciences, Siberian Branch, 660036, Krasnoyarsk, Russia; rsa@iph.krasn.ru*

We calculate the time evolution of mean spin components and the squared I-concurrence of two coupled large spins S. As the initial conditions we take two cases: spin coherent states and uniform superposition states. For the spin coherent states we have obtained the asymptotic formulas at S>>1 and t<<T (T is period). We draw a conclusion that quantum computation on qudits, represented by large spins, do not get the principal limitations on the spin number S.

## Introduction

In recent years, quantum computation not only on quantum systems with two levels (qubits) but on d-level systems (qudits) [1-3] is investigated. Qudit quantum information processing employs fewer coupled quantum systems: a considerable advantage for the experimental realization of quantum computing:

$$n = \ln N / \ln d \quad \Leftrightarrow \quad n = \ln N / \ln 2 .$$

A particle with large spin S, for example a quadrupole nucleus or a single-molecule magnet, can be treated as a qudit with d=2S+1. Interactions of the spin with a static magnetic field and with an electric field gradient result in formation non-equidistant energy levels, allowing to selective control of the states. We can ask: is there limitation on the value of S? Will quantum behaviour (superposition and entanglement) be retained with increasing in S?

We can found papers (see e.g. [4-5]), in which it is expected that with increasing of S will be the transition from quantum to classical spin angular momentum. However, another authors [7-8] do not share this expectation, because the entanglement [8-9] and the boundary of Bell's inequalities [7] increase with increasing S. To study this problem, we now consider the quantum dynamics of a simple system: two coupled large spins. We calculate the time evolution of spin components and the squared I-concurrence [10], which is an entanglement measure.

## The system

The spin-spin interaction between two spins S is given by the Ising-like Hamiltonian

$$H_{SS} = -\frac{J}{S} S_1^Z S_2^Z , \qquad (1)$$

where $S_i^Z$ is Z component of spin-S angular-momentum operator $\vec{S}_i$ (i=1, 2). To study the limit S $\to \infty$, we have scaled a constant interaction inversely with S, follow B.C. Sanders [4].

Let us use the basis of eigenstates of $S_i^Z$ (with certain value at the projection on Z-axis for each spin): $|m_1\rangle \otimes |m_2\rangle$, where $m_1$ and $m_2$ are 2S +1 values: -S, -S+1, . . . , S-1, S.

Suppose that the two spins are initially in a pure product state:

$$|\Psi\rangle = |\Psi_1\rangle \otimes |\Psi_2\rangle, \qquad |\Psi_i\rangle = \sum_{m=-S}^{m=S} C_m |m\rangle,$$

where $|\Psi_i\rangle$ is an arbitrary superposition state. This state under the action of the Hamiltonian (1) becomes time dependence

$$|\Psi(t)\rangle = \sum_{n=-S}^{n=S} \sum_{m=-S}^{m=S} C_n C_m \exp\left\{itmn\frac{J}{S}\right\} |m\rangle \otimes |n\rangle. \qquad (2)$$

Note that due to the discreteness of the levels the quantum dynamics of the isolated system is periodic with a recurrence time of $T = \frac{4\pi S}{J}$.

For concrete calculations, we choose two important variants of the initial conditions:
1) the spin coherent states - eigenstates of $S_i^X$ with maximal X-projection [9,11]:

$$|S\rangle_X = \frac{1}{2^S} \sum_{m=-S}^{m=S} \binom{2S}{S+m}^{1/2} |m\rangle_Z; \qquad (3)$$

2) the uniform superposition states [12]:

$$|\Psi_S\rangle = \frac{1}{d^{1/2}} \sum_{m=-S}^{m=S} |m\rangle. \qquad (4)$$

These states can be prepared from the ground state $|\Psi_0\rangle = |S\rangle_Z \otimes |S\rangle_Z$ by following operations to both spins for each of the two variants, respectively:
1) π/2-rotation around the Y-axis (the nonselective operation);
2) the operator of the quantum Fourier transform QFT$_d$ (we found the sequence of selective rotations for d=3÷10 [13]).

**The motion of spin components**

Calculate the time evolution of the mean value of the first spin X-projection $\langle S_1^X(t) \rangle$. As usually, the spin operator can be decomposed into the ladder operators:

$$S^X = (S^+ + S^-)\frac{1}{2},$$

which have only nonzero matrix elements between neighboring states

$$\langle m|S^+|m-1\rangle = \langle m-1|S^-|m\rangle = (S-m+1)^{1/2}(S+m)^{1/2}.$$

For function normalized to 1 at t = 0, we obtain time dependence

$$F(t) = \frac{\langle \Psi(t)|S_1^+|\Psi(t)\rangle}{\langle \Psi(0)|S_1^+|\Psi(0)\rangle} = \sum_{n=-S}^{n=S} |C_n|^2 \exp\left\{itn\frac{J}{S}\right\}, \qquad (5)$$

$$\langle\Psi(0)|S_1^+|\Psi(0)\rangle = \sum_{m=-S+1}^{m=S} C_m^* C_{m-1}(S-m+1)^{1/2}(S+m)^{1/2}.$$

For the two variants, we have, respectively:

1) $$F_{coh}(t) = \left(\cos\frac{Jt}{2S}\right)^{2S}, \quad \langle\Psi(0)|S_1^+|\Psi(0)\rangle = S; \tag{6}$$

2) $$F_{sup} = \frac{\sin(Jt(1+1/(2S)))}{(2S+1)\sin(Jt/(2S))}, \quad \langle\Psi(0)|S_1^+|\Psi(0)\rangle = \frac{1}{2S+1}\sum_{m=-S+1}^{m=S}(S-m+1)^{1/2}(S+m)^{1/2}.$$

The dependencies are shown in Fig. 1 for small (a) and large (b) spin number.

The period of oscillation is increasing with the growth of S. When S >> 1, t << T we can approximate the functions as follows: by Gaussian functions at first and by sine divided by Jt at second -

$$F_{sup} = \frac{\sin(Jt)}{Jt}, \quad F_{coh}(t) \cong F_g(t) = \exp\left(-\frac{(Jt)^2}{4S}\right). \tag{7}$$

In the limit $S \to \infty$ for the spin coherent states, we get a constant, i.e. the result coincides with the result for two classical magnetic moments, lying in the plane, because the interaction (1) turns to zero as $S_1^Z = 0, S_2^Z = 0$. The second state keeps a superposition which non-represented classically.

**Squared I-concurrence**

As an entanglement measure we will use the squared I-concurrence [10], that for pure states takes the form,

$$C^2 = \frac{d}{d-1}\left(1 - Tr_1\hat{\rho}_1^2\right), \tag{8}$$

where $\hat{\rho}_1 = Tr_2|\Psi(t)\rangle\langle\Psi(t)|$ is the reduced density matrix of the first spin (traced by the second spin), which we find

$$\hat{\rho}_1 = \sum_{m_1=-S}^{m_1=S}\sum_{m_2=-S}^{m_2=S} C_{m_1}C_{m_2}^*|m_1\rangle\langle m_2|\sum_{n=-S}^{n=S}|C_n|^2\exp\left\{itn(m_1-m_2)\frac{J}{S}\right\}.$$

Let us compare this formula with the previous case (5). There transitions between near levels $|m_1 - m_2| = 1$ take part only. Now all transitions take part and result in double sum and large coefficient $|m_1 - m_2| \approx S$ in exponent. That is mathematics basis of differences in time behaviors of the mean spin component and the entanglement.

Calculation gives:

1) The coherent state:

$$Tr_1\hat{\rho}_1^2 = \frac{2}{2^{4S}} \sum_{M=1}^{M=2S} \binom{4S}{2S+M}\left[\cos\left\{M\frac{Jt}{2S}\right\}\right]^{4S} + \frac{1}{2^{4S}}\binom{4S}{2S}, \qquad M=(m_1-m_2). \quad (9)$$

At small times $\quad C_{coh}^2 = \frac{2S+1}{4S}(Jt)^2 + O(t^4).$

The time average for the period $\quad \overline{C}_{coh}^2 = \frac{2S+1}{2S}\left(1-(2)^{-4S}\binom{4S}{2S}\right)^2 \cong \frac{2S+1}{2S}\left(1-\frac{1}{\sqrt{2\pi S}}\right)^2$

2) The uniform superposition state:

$$Tr_1\hat{\rho}_1^2 = \frac{2}{d^4}\sum_{M=1}^{d-1}(d-M)\frac{1-\cos(JtMd/S)}{1-\cos(JtM/S)} + \frac{1}{d}. \quad (10)$$

At small times $\quad C_{sup}^2 = \frac{(d+1)d^3}{(12S)^2}(Jt)^2 + O(t^4) \approx \frac{S^2}{9}(Jt)^2.$

The time average for the period $\quad \overline{C}_{sup}^2 = 1 - \frac{1}{d} = \frac{2S}{2S+1}.$

In Refs. [9, 11, 12] have been studied the entanglement for such states. The results are similar.

The time dependences of the squared I-concurrence in some values of S, calculated by the obtained formulas, are shown in Fig. 2. We may see that when the spin numbers increase, the squared I-concurrence grows and is closed in 1. This deviation from 1 for the uniform superposition state is less than for the coherent state. There are the time oscillations. Starting with the time T/4 the entanglement will be decreased. G. Burlak, I. Sainz, and A.B. Klimov [9] found that instead of decrease can be obtained a further increase in the concurrence if to apply $\pm\pi/2$ rotations around the Y-axis to both spins at some appropriate time moments: $t_1$ and $t_2$.

**Asymptotic formula for the squared I-concurrence of coherent states**

In the case of coherent states with S>>1 for the squared I-concurrence can be obtained a simple asymptotic formula. In Eq. (9) for $Tr_1\hat{\rho}_1^2$ when t<<T, we replace approximately the binominal coefficients and the cosine to the high power 4S by Gaussian functions:

$$\frac{1}{2^{4S}}\binom{4S}{2S+M} \approx \frac{1}{\sqrt{2\pi S}}\exp\left(-\frac{M^2}{2S}\right),$$

$$\left[\cos\left\{M\frac{Jt}{2S}\right\}\right]^{4S} \approx \exp\left\{-M^2\frac{(Jt)^2}{2S}\right\},$$

and summation by integration. Then we obtain

$$Tr_1\hat{\rho}_1^2 \approx \frac{1}{\sqrt{2\pi S}} \int_{-2S}^{2S} \exp\left\{-\frac{M^2}{2S}(1+(Jt)^2)\right\} dM - \frac{1}{(Jt)^2+1} erf\sqrt{\frac{J^2t^2+1}{8S}} + \frac{1}{\sqrt{2\pi S}}. \quad (11)$$

Here we took into account the special role of summands $M = (m_1 - m_2) = 0$, which do not vary over time and, therefore, does not decay. Their contribution is

$$Tr_1\hat{\rho}_1^2 = (2)^{-4S}\binom{4S}{2S} \cong \frac{1}{\sqrt{2\pi S}}.$$

Let us replace approximately 2S by ∞ in the integration limits, and then we find

$$C_{coh}^2 = \frac{2S+1}{2S}\left(1 - \frac{1}{\sqrt{1+J^2t^2}}(1 - erf\sqrt{\frac{J^2t^2+1}{8S}}) - \frac{1}{\sqrt{2\pi S}}\right). \quad (12)$$

Consider again the exact time dependence of squared I-concurrence in Fig.2. With increased time on the background of monotonous growth we see a sequence of local minima in the points $t\frac{JM}{2S} = n\pi$ in which the cosine is drawn to ± 1. When S>>1, these minima can be described by Gaussian functions as:

$$\left[\cos\left\{M\frac{Jt}{2S}\right\}\right]^{4S} \approx \sum_{n=0}^{n=M} \exp\left\{-\frac{M^2}{2S}J^2\left(t - \frac{nT}{2M}\right)^2\right\}.$$

Then to get a squared I-concurrence

$$C_{coh}^2 = \frac{2S+1}{2S}\left(1 - \frac{1}{\sqrt{1+J^2t^2}}(1 - erf\sqrt{\frac{J^2t^2+1}{8S}}) - \frac{1}{\sqrt{2\pi S}} - \frac{2}{\sqrt{2\pi S}}\sum_{M=1}^{M=2S}\sum_{n=1}^{n=M}\exp\left\{-\frac{M^2}{2S}\left[1 + J^2\left(t - \frac{nT}{2M}\right)^2\right]\right\}\right). \quad (13)$$

In Fig. 3 performed a comparison of the dependences, calculated on the exact and asymptotic formulas. In the last (13), due to the rapid decrease of amplitude with increasing M, we are restricted by the first three ingredients with M = 2, 3 and 4. We see good agreement the asymptotic formula with exact one. When S increases its accuracy will increase, and the depth of minima will be reduced.

The changes in behaviors $F_{coh}(t)$ and $C_{coh}^2(t)$ with the growth of S shown in Fig. 4. The time evolution of $F_{coh}(t)$ slows down and thus closes in the motion of classical magnetic moments. While the squared I-concurrence reaches maximum value C = 1, indicating preservation of the quantum properties. Previously, such a dependence of entanglement has been demon-strated in another model system: a particle with spin, oscillating in a non-uniform magnetic field [8].

**Conclusions**

If the two spins are prepared in the spin coherent states, the motion of their mean projections in the limit $S \to \infty$ will concede with the motion of the projections of two classical magnetic moments. Nevertheless, the squared I-concurrence, and thus the entanglement will grow with time to values approaching to the maximum $C = 1$ when $S \to \infty$.

If the two spins are prepared in the uniform superposition states, we observe the quantum dynamics in any S, so far as classical magnetic moments can not be in superposition state (like of the famous «Schrodinger cat» [14]). (Although the preparation of the superposition state at large S can be technically very difficult).

Thus, the large spins may show both the quantum and classical properties, according to prepared states and observe conditions. It is important for us (and this is the main goal of this study) that quantum computation on qudits, represented by large spins, do not get the principal limitations on the spin number S.

This work was supported by the Russian Foundation for Basic Research (project. no. 09-07-00138-a).

# Квантовая динамика двух взаимодействующих больших спинов

Владимир Евгеньевич Зобов

*Институт физики им. Л. В. Киренского СО РАН,*

*660036, Красноярск, Академгородок,50, стр.38.*

*rsa@iph.krasn.ru*

Рассчитывается временная эволюция средних спиновых компонент и квадратичной l-согласованности двух взаимодействующих больших спинов $S$. В качестве начальных условий взяты два случая: спиновые когерентные состояния и равномерные суперпозиционные состояния. Для спиновых когерентных состояний получено асимптотическое выражение при $S \gg 1$ и $t \ll T$ ($T$-период). Сделан вывод, что квантовые вычисления на кудитах, представленных большими спинами, не встречают принципиальных ограничений на спиновое число $S$.

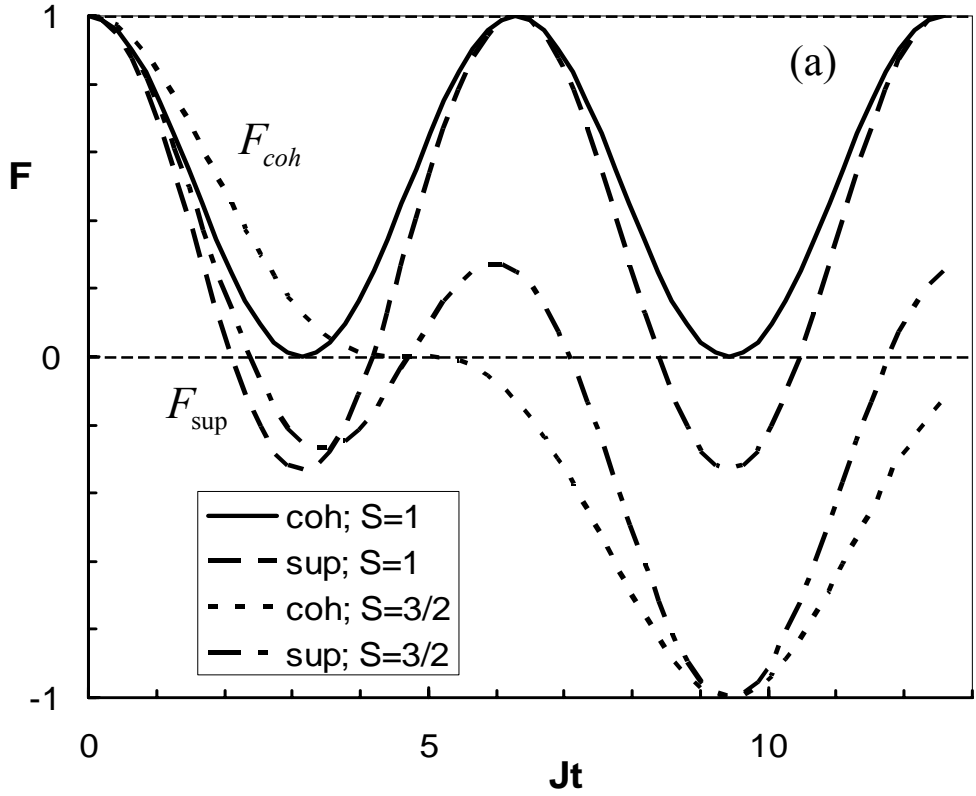

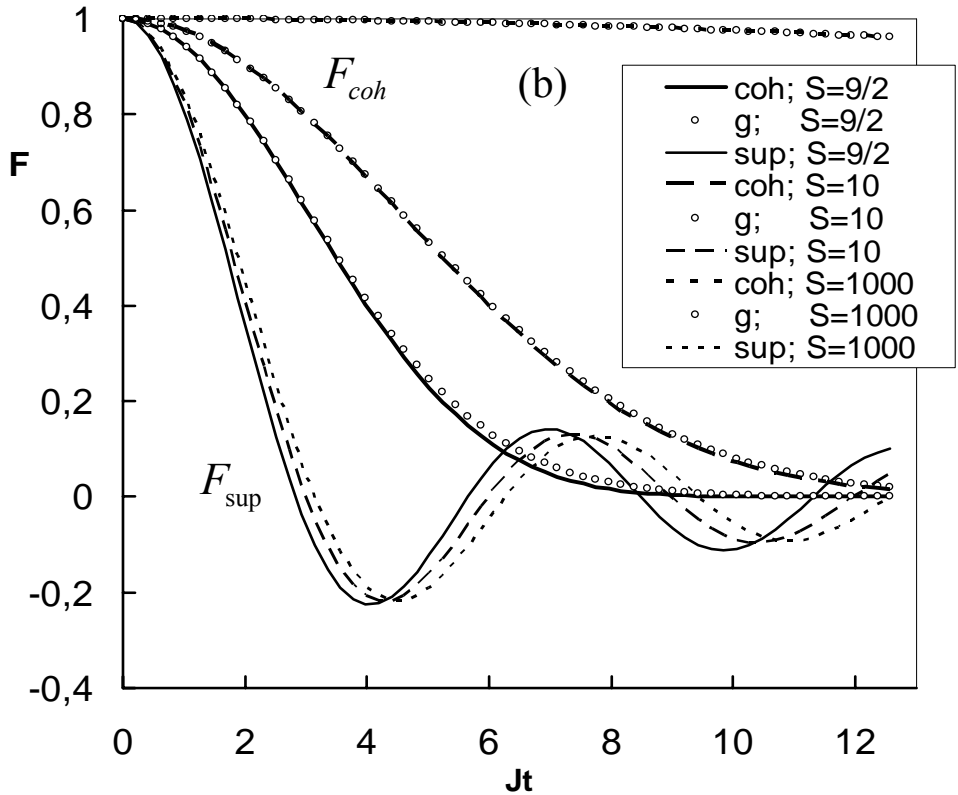

Fig. 1. Time evolution of the normalized mean value of the first spin X-projection for small S (a) and large S (b) (the Gaussian functions $F_g(t)$ (7) for correspondence S are shown by open circles).

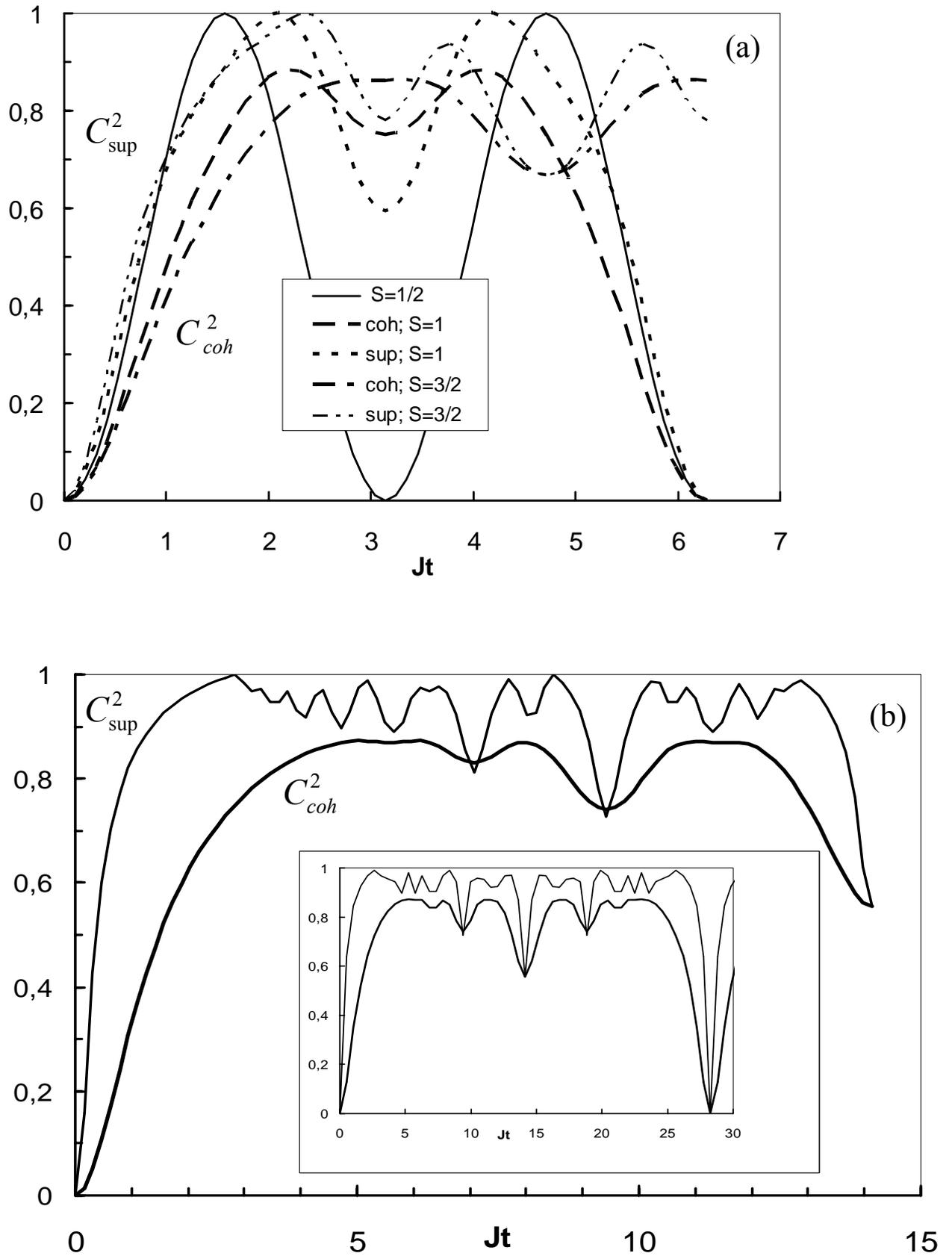

Fig. 2. Time evolution of the squared I-concurrence for different values of S: ½, 1, 3/2 (a), and 9/2 (b) where inset shows this dependence up to time T/2.

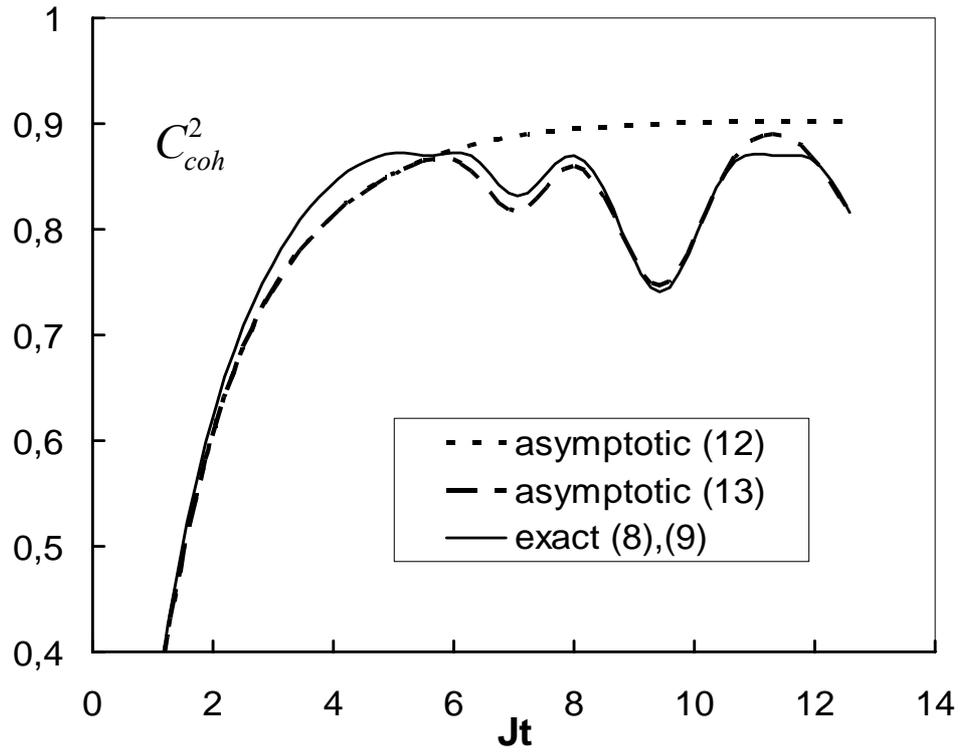

Fig. 3. Time evolution of the squared I-concurrence in the spin coherent states for S = 9/2, calculated on the exact formula (8) with (9) and the asymptotic formulas (12) and (13).

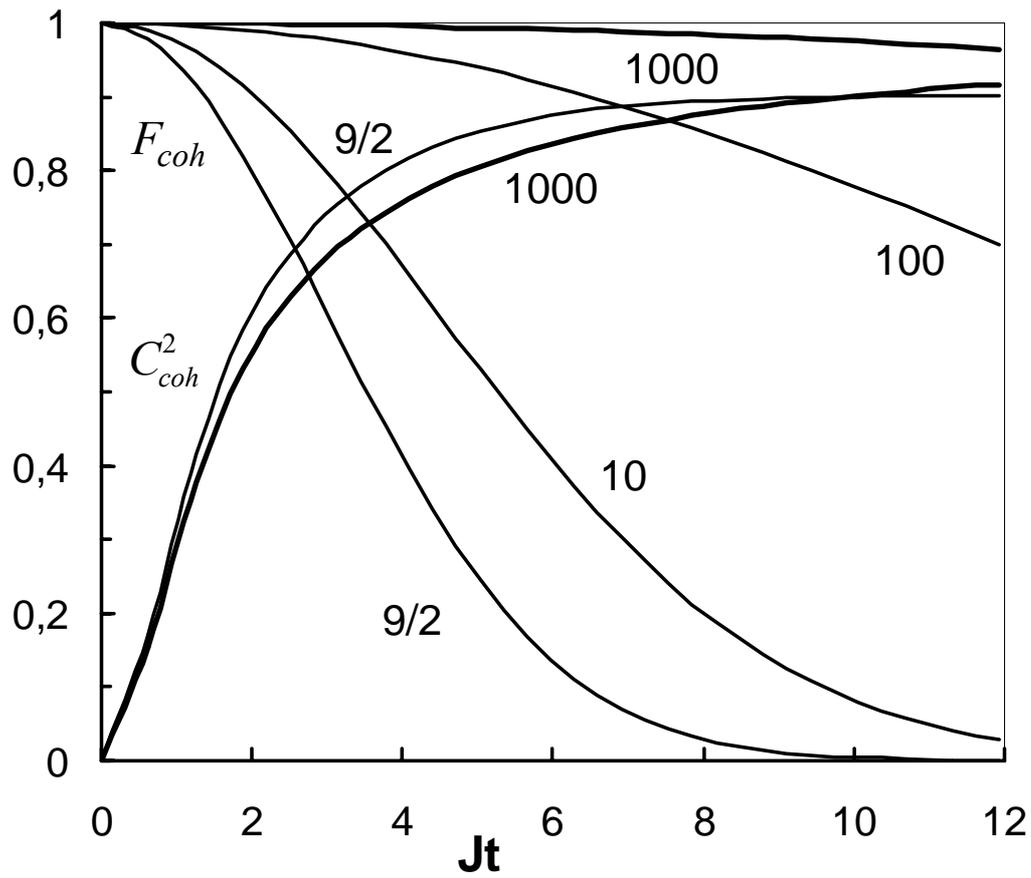

Fig. 4. Time evolution of $F_{coh}(t)$ and $C^2_{coh}(t)$, for different values of S (the numbers on the curves), calculated on the asymptotic formulas (7) and (12), respectively.